# Energetic sub-2-cycle laser with 220 W average power

Steffen Hädrich,[1,2]* Marco Kienel,[1,2] michael müller,[1] Arno Klenke,[1,2] Jan Rothhardt,[1,2] Robert klas,[1,2] thomas gottschall,[1] tino eidam,[3] andrás drozdy,[4] péter jójárt,[4] zoltán várallyay,[4] eric cormier,[4,5] károly osvay,[4] andreas tünnermann,[1,2,6] and jens limpert[1,2,3,6]

[1]Institute of Applied Physics, Abbe Center of Photonics, Friedrich-Schiller-Universität Jena, Albert-Einstein-Straße 15, 07745 Jena, Germany
[2]Helmholtz Institute Jena, Fröbelstieg 3, 07743 Jena, Germany
[3]Active Fiber Systems GmbH, Wildenbruchstr. 15, 07745 Jena, Germany
[4]ELI-ALPS, ELI-HU Non-Profit Ltd., H-6720 Szeged, Dugonics tér 13, Hungary
[5]Université Bordeaux-CNRS-CEA-UMR 5107, CELIA, 351 Cours de la Libération, F-33405 Talence, France
[6]Fraunhofer Institute for Applied Optics and Precision Engineering, Albert-Einstein-Straße 7, 07745 Jena. Germany
*Corresponding author: steffen.haedrich@uni-jena.de

**Few-cycle lasers are essential for many research areas such as attosecond physics that promises to address fundamental questions in science and technology. Therefore, further advancements are connected to significant progress in the underlying laser technology. Here, two-stage nonlinear compression of a 660 W femtosecond fiber laser system is utilized to achieve unprecedented average power levels of energetic ultrashort or even few-cycle laser pulses. In a first compression step 408 W, 320 µJ, 30fs pulses are achieved, which can be further compressed to 216 W, 170 µJ, 6.3 fs pulses in a second compression stage. This is the highest average power few-cycle laser system presented so far. It is expected to significantly advance the fields of high harmonic generation and attosecond science.** © 2016 Optical Society of America under Open Access Publishing Agreement. Users may use, reuse, and build upon the article, or use the article for text or data mining, so long as such uses are for non-commercial purposes and appropriate attribution is maintained. All other rights are reserved.

http://dx.doi.org/10.1364/OL.41.004332

Intense ultrashort and few-cycle pulses have become available and rapidly found numerous applications in science and technology [1]. Their short pulse duration, for example, allows to study ultrafast processes in atoms, molecules and more complex physical systems [2]. Another very prominent example is the use of such lasers to generate coherent extreme ultraviolet (XUV) radiation via the process of high harmonic generation (HHG) [3,4]. Since HHG can be achieved with a table-top laser system these XUV sources are particularly interesting as a complementary source to large-scale facilities such as synchrotrons and free-electron lasers. Consequently, they have found numerous applications in atomic- and molecular physics [5,6], material science [7–9], imaging techniques [10] and many others [11]. Furthermore, the process of HHG is inherently linked to the generation of attosecond pulses allowing to study even faster processes, i.e. on the time scale of electronic motion, than accessible with the ultrashort infrared driving pulses alone [2]. In gas phase, attosecond pulses are emitted during each half-cycle of the driving electrical field. When the pulse duration of the laser pulse is reduced to less than two cycles in duration and the electrical waveform can be controlled, i.e. a stabilized carrier envelope phase (CEP) is achieved, an isolated attosecond pulse is generated when spectrally filtering the cutoff harmonics [12]. With the availability of CEP stabilized few-cycle lasers attosecond science has emerged as a complete new research field, which addresses many fundamental questions in physics, chemistry, biology, medicine, material science and many others. Despite impressive progress in the afore-mentioned field not only significant advancements in laser technology are required to push the frontiers, but this technology has to be made available to a large user community. This is elucidated by the mission goals of the Extreme Light Infrastructure Attosecond Light Pulse Source (ELI-ALPS), which is currently under construction and aiming to provide a unique source of attosecond pulses for users. One of the beam lines, the high repetition rate (HR) laser system that will be operated, requires a 1 mJ, sub-2 cycle laser running at 100 kHz repetition rate (100 W average power) [13], which is well beyond the current state-of-the-art.

Intense few-cycle laser pulses are mainly achieved via optical parametric chirped pulse amplification (OPCPA) or nonlinear compression in gas-filled capillaries. OPCPA relies on nonlinear interaction of a strong pump laser and a weak broadband signal to amplify the latter one [14]. Commonly used nonlinear crystals allow for an extremely broad amplification bandwidth as required for few-cycle pulses [15,16]. The pump laser dictates the performance of an OPCPA system in terms of achievable pulse energy, repetition rate and average power. Advancements in picosecond pump laser technology have made available kilowatt level pump lasers that in principle should allow for high average power OPCPA systems [17]. However, the average power of such systems has not exceeded 22 W so far [18]. This limitation is mainly due to linear absorption in the nonlinear crystals inducing thermo-optical effects or eventually damaging the crystals via heat induced tension [19]. There has been effort to mitigate this effect, e.g. by sandwiching the crystals with highly conducting materials [20], but further average power scaling is severely challenged by the thermal effects. Nonlinear compression, on the other hand, has been the standard approach to compress 20-30 fs Ti:Sapphire lasers to sub-5 fs and is considered the workhorse for attosecond science so far [21,22]. However, the employed laser technology has limited the repetition rate and average power to a few kHz and Watt, respectively. In recent years, fiber lasers have successfully been used to significantly increase both of these parameters and have even achieved few-cycle pulses [23,24]. Furthermore, tests with kilowatt continuous wave (CW) lasers promised excellent power-scaling capabilities of this approach [25], which makes it a viable alternative to OPCPA systems.

Here, we present results on two-stage nonlinear compression of a 660 W, 0.52 mJ, sub-300 fs fiber laser systems allowing to achieve the highest average power few-cycle laser system to date. It provides 216 W, 170 µJ, 6.3 fs (sub-2 cycle) pulses, which can be employed for a great variety of applications.

Figure 1 shows the schematic of the setup used for nonlinear compression. The frontend is a fiber chirped pulse amplification (FCPA) system incorporating coherent combination of up to 8 main amplifier channels. A detailed description of the laser system and the addressable laser pulse parameter range can be found in [26]. For the experiments presented here this system delivers up to 660 W of average power at a repetition rate of 1.27 MHz corresponding to a pulse energy of 520 µJ. The compressed pulse duration is 240 fs leading to a peak power of approximately 2 GW.

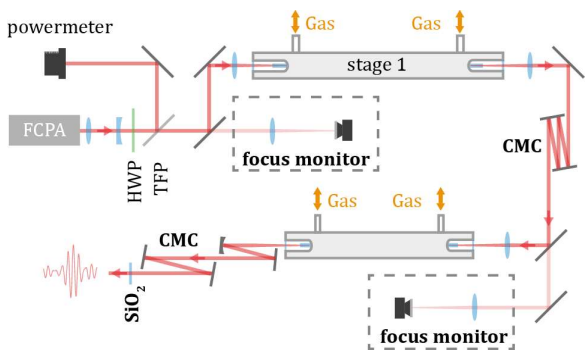

**Fig. 1.** Generic setup of the two-stage nonlinear compression experiment. (FCPA – Fiber chirped pulse amplification system, HWP – half-wave plate, TFP – Thin film polarizer, CMC – chirped mirror compressor)

As shown in Fig. 1 the laser beam traverses a telescope for adaption of the spatial beam size and a combination of half-wave plate (HWP) and thin-film polarizer (TFP). This combination allows to arbitrarily change the laser power transmitted through the TFP to the nonlinear compression experiment. A lens is used to couple the laser beam to a 1 m long capillary with an inner diameter of 250 µm, which sits on a water-cooled V-groove in a pressure chamber [25]. In a first step the whole laser power is directed towards the power meter by appropriately turning the HWP to measure the average power. Then, a small fraction is sent to the experiment to optimize the coupling. A focus monitor installed into the leakage of one of the input beam steering mirrors (Fig.1) allows to precisely adapt the focal spot size at the entrance of the capillary in stage 1. This is crucial to achieve the best possible coupling and highest transmission efficiency through the capillary. Subsequently, the average power is gradually increased up to the full available value of 660 W. When filling the capillary with argon gas at a pressure of 3.5 bar significant spectral broadening is achieved via self-phase modulation in the noble gas filled capillary as shown in Fig. 2a). Temporal compression is achieved with a chirped mirror compressor (CMC) that has a group delay dispersion (GDD) of -1400 fs$^2$. Due to the efficient coupling the average power after the CMC is as high as 408 W corresponding to a pulse energy of 320 µJ. Figure 2b) shows the autocorrelation of the compressed pulses, which indicates a pulse duration of around 30 fs.

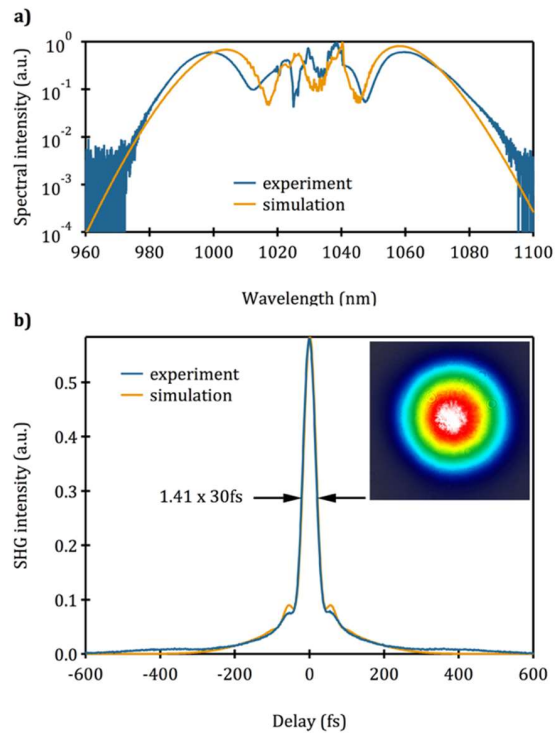

**Fig. 2.** Experimental results of the first nonlinear compression stage. Propagating the FCPA pulses in a capillary filled with 3.5 bar of argon gas leads to spectral broadening as shown in the blue curve in a). An autocorrelation of the compressed pulses is shown in the blue curve in b). The inset in b) shows the intensity profile of the collimated beam measured after the chirped mirror compressor. The orange curve in a) and b) correspond to a numerical simulation of the compression experiment (see text for details).

Owing to the spatial filtering occurring during the propagation in the capillary, the beam profile emerging from the first stage is excellent (inset of Fig.2b). Although not measured here, previous experiments have already shown that beam quality factors on the order of M²=1.1 can be obtained when using capillaries [27]. Additionally, the process of spectral broadening and subsequent temporal compression is simulated using a commercially available tool based on a split-step Fourier transform method to solve the extended nonlinear Schrödinger equation [28]. The simulation includes self-phase modulation, self-steepening and four-wave mixing processes, but no dispersion, since it is negligible for the bandwidth in the first stage. The input pulses for the simulation are obtained from a measurement of the spectrum and the spectral phase of the FCPA system, which is not shown here. The results are in good agreement with the measured data both for the spectrum and the autocorrelation as indicated by the orange curves in Fig. 2. Please note that in the simulation the nonlinear refractive index $n_2$ is adapted so as to match the measured broadened spectrum. The good agreement with the autocorrelation trace strengthen the deduced pulse duration of 30 fs. From the temporal pulse shape simulation a peak power of 7.7 GW is obtained. This high peak power ultrashort laser pulses already constitute an average power record and could be readily applied to frequency conversion into the XUV spectral range via HHG [29].

Further reduction in pulse duration requires a second compression stage as shown in Fig.1. The pulses emerging from the first compression stage are coupled into a second capillary with the same inner diameter of 250 µm, but with a shorter length of ~0.6 m. Due to the large spectral bandwidth all the optics after the exit of the second capillary are reflective. After passing through a thin anti-reflection coated window the beam is collimated with a spherical mirror (f=500 mm) that is used in combination with a flat mirror (Fig.1). These mirrors are specially designed dielectric mirrors to steer high average power few-cycle pulses. They not only exhibit a high reflectivity (730-1250 nm) over a broad spectral range, but they are also designed to have a flat GDD across the same bandwidth. This is achieved by using mirror pairs, in that case the spherical and the flat mirror, that have opposite GDD that compensates each other similar to the concept applied to double-angle chirped mirrors. The major benefit is that they have already shown excellent average power capability compared to conventional metal mirrors [25], which have shown thermal effects at average power levels of 30 W [24]. The final temporal compression is achieved by two reflections on broadband chirped mirrors with a total GDD of –100 fs² and a 1 mm piece of anti-reflection coated fused silica substrate (GDD of 19 fs²). When filling the second capillary with 7.5 bar of neon gas additional spectral broadening covering the range between 700 nm and 1250 nm (blue curve in Fig. 3a)) is achieved. The spectral dip at approximately 720 nm is due to a transmission dip of the GDD optimized broadband mirrors. After the second CMC a small fraction of the beam is sent to a SPIDER device to measure the spectral phase (red curve in Fig 3 a). This allows to obtain the pulse profile shown in Fig. 3b) that indicates a pulse duration of only 6.3 fs, which is less than 2 cycles at 1030 nm and significantly shorter than our previous demonstration [24]. In addition, the average power is as high as 216 W after the compressor corresponding to a pulse energy of 170 µJ at 1.27 MHz. This few-cycle laser system delivers about one order of magnitude higher average power as compared to OPCPA systems [18] and a factor of four more than previous demonstrations with hollow fiber compressors [24]. Although, the SPIDER measurement allows for a measurement of the spectral phase an additional simulation is performed to gain more information on the temporal pulse profile. For that purpose, the simulation output from stage 1 (orange curves in Fig. 2) is used as input for the second stage. The calculation is done similar to the one before, but here with the dispersion of gas-filled capillary included. Following the model introduced by Travers et al. [30] the following dispersion terms are deduced for a 250µm capillary filled with 7.5 bar of neon: $\beta_2$=-3.4 fs²/m, $\beta_3$=6.0 fs³/m and $\beta_4$=-2.7 fs⁴/m for 1030 nm central wavelength. Furthermore, the $n_2$ value is adapted to qualitatively reproduce the spectral broadening (orange curve in Fig. 3 a)). Compared to the simulation performed for stage 1, there is a slight deviation in the spectral shape at long wavelengths, which could be caused by the measurement itself or by a slightly different temporal pulse shape from stage 1 (the stage 2 calculation is seeded by the stage 1 simulation not the real pulse). The latter one affects the final spectral shape through the process of SPM significantly [31].

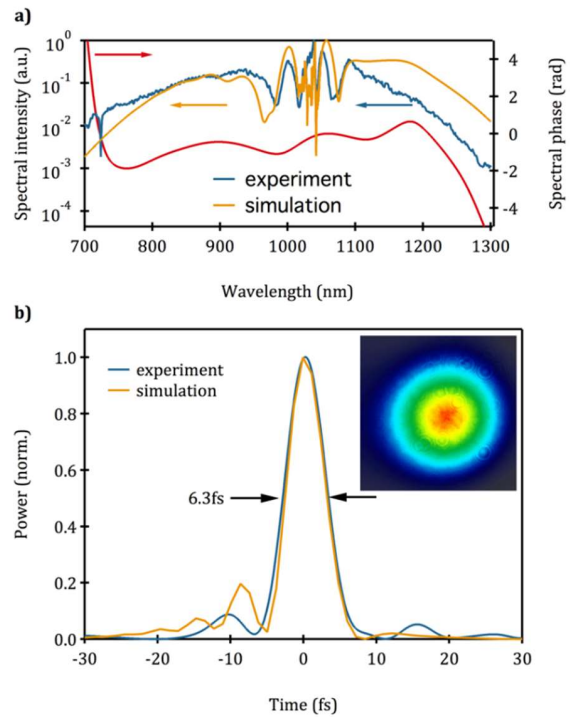

**Fig. 3.** Experimental results of the second nonlinear compression stage. The compressed output from stage 1 (Fig. 2) is propagated in a second capillary filled with 7.5 bar of neon gas to achieve further spectral broadening as shown in the blue curve in a). A SPIDER measurement is used for temporal pulse characterization (blue curve in b)). The inset in b) shows the collimated beam profile measured after the CMC. The orange curve in a) and b) correspond to a numerical simulation of the compression experiment (see text for details).

The compressed pulse shapes of the simulation (orange curve in Fig. 3b)) and the experiment (blue curve in Fig. 3b)) agree fairly well in the main part of the pulse both leading to a pulse duration of 6.3 fs. The experiment seems to lack the missing pre- and post-pulses predicted by the model. This can be attributed to rapid phase modulations occurring around the central part (~1030 nm) that are not resolved by the SPIDER measurement leading to further side pulses. Since the SPIDER measurement seems to overestimate the actual energy in the main peak, the temporal waveform given by the simulation is used to estimate a minimal peak power as high as 17 GW in this case. It has to be noted that the pulse compression can be significantly improved by employing customized chirped mirrors to compensate for higher order phase terms in both compression stages. The beam profile is excellent as indicated by the inset in Fig. 3b. However, it has to be noted that this profile only accounts for spectral content below 1100 nm, since it has been measured with a silicon-based CCD camera. Consequently, the system delivers a high average power sub-2 cycle pulses with a high peak power and good beam quality.

In summary, two-stage nonlinear compression of a 660 W femtosecond FCPA system is successfully demonstrated. Combining this high average power laser system with a power scalable compression approach allows for unprecedented average levels for ultrashort to few-cycle laser systems. The first compression stage delivers 408 W, 30 fs with a pulse energy of 320 µJ. Similarly compressed pulses from FCPA systems have recently been successfully used for a number of experiments in HHG and subsequent applications [29,32,33]. Therefore, it can be expected that the here presented system can readily boost the available photon flux in HHG up to one order of magnitude paving the way for novel experiments, e.g. in time-resolved coincidence studies of inner-shell ionization processes in molecules [32], photoemission spectroscopy [9,34] or nano-scale imaging [10,32]. The subsequent compression to sub-2 cycle pulses, 170 µJ pulses with 17 GW peak power and 216 W of average power constitutes a significant increase as compared to OPCPA [18] and previous nonlinear compression experiments [24]. This opens up a plethora of new research possibilities. For example, such a laser system could be used for high photon flux soft x-ray generation up to the water window [24]. Additionally, the validation of nonlinear compression as a power scalable concept for few-cycle pulse generation is an important step for realizing the next generation sources as envisioned by the ELI-ALPS facility. Regarding the latter, future work will focus on increasing average power as well as energy scaling to achieve similar performance parameters at up to 1 mJ of pulse energy and beyond after the second compression stage. Obviously, the need for carrier envelope phase stabilization has to be addressed in upcoming work, finally, unraveling the full potential of such high average power energetic few-cycle laser systems.

**Funding.** This project has received funding from the ELI-HU Non Profit LTD under contract number "NLO3.7 Uni of Jena" and the European Research Council (ERC) under the European Union's Horizon 2020 research and innovation program (grant agreement No [670557], MIMAS).

**Acknowledgment**. We thank S. Demmler for fruitful discussions regarding the SPIDER measurement and T. Gross of Laseroptik GmbH for the GDD mirror design.